\documentclass[11pt]{article}

\usepackage{amssymb,amsmath,latexsym}
\usepackage{amsfonts}
\usepackage{enumerate}
\usepackage{multirow,xcolor,colortbl}
\usepackage {color}
\usepackage[pdftex] {graphicx}
\usepackage[left=2cm,right=2cm,top=4cm,bottom=4cm,a4paper]{geometry}
\usepackage{here}
\usepackage{amsthm}
\usepackage{breqn}
\usepackage{relsize}
\usepackage{subfigure}
\usepackage{array,arydshln} 
\usepackage{hhline} 
\usepackage{lmodern}
\usepackage{manyfoot}
\usepackage{booktabs}
\usepackage{xcolor}

\usepackage[utf8x]{inputenc}
\usepackage{amssymb,amsmath,latexsym}
\usepackage{amsfonts}
\usepackage{enumerate}
\usepackage{multirow,xcolor,colortbl}

\usepackage[english]{babel}
\usepackage {color}
\usepackage[pdftex] {graphicx}
\usepackage[left=2cm,right=2cm,top=4cm,bottom=4cm,a4paper]{geometry}
\usepackage{here}
\usepackage{amsthm}
\usepackage{breqn}
\usepackage{relsize}
\usepackage{subfigure}
\usepackage{array,arydshln} 
\usepackage{hhline} 
\usepackage{lmodern}
\usepackage{manyfoot}
\usepackage{booktabs}
\usepackage{xcolor}

\begin{document}
%
\title{Two Segmentation Methods for the Diagnosis of Malignant Melanoma}
%
%
%

\author{Seungmin~Park\dag, ~Hyunju~Lee\dag, ~and~Kiwoon~Kwon\dag\ddag     
\thanks{\dag Department of Mathematics, Dongguk University,     \ddag  e-mail:kwkwon@dongguk.edu}}

\maketitle

\begin{abstract}
Automatic diagnosis of malignant melanoma highly depends on the segmentation methods used for the suspicious lesion.
We suggest the parameter selection method (PSM) and maximum area method (MAM) for the segmentation of the lesion to be diagnosed.
Herein, these segmentation methods are compared to a skin cancer expert's segmentation and three other conventional algorithms.
The diagnosis of malignant melanoma based on the two suggested, three conventional, and expert's segmentation  are compared with respect to sensitivity, specificity, and accuracy.
\end{abstract}


%

\section{Introduction}
In the USA, the melanoma incidence rate has increased over the past 30 years. Although melanoma constitutes about 1$\%$ of all skin cancers,  a large proportion of  skin cancer-related deaths come from melanoma since its metastasis is very fast.  The American cancer society estimate there will be 100,350 diagnosis and 6,850 deaths from melanoma in 2020 \cite{Cancer2020}.  Early diagnosis of malignant melanoma is very important to prevent melanoma from metastasizing to other sites in the body. An automated system might facilitate the early detection of malignant melanoma, since even experienced dermatologists experience fatigue and can have difficulty in diagnosing malignant melanoma  \cite{Overview, Overview2, Overview3, BMEL1}.

Melanoma diagnosis is usually composed of three important steps: Segmentation of the lesion, Feature extraction from the segmented lesion, and Classification of malignant from benign melanomas based on the extracted features \cite{Oludayo, Mehta, Masood}. The following methods 
will be used in these three major, along with a few minor, steps to diagnose malignant from benign melanomas:
\begin{itemize}
\item{Preprocessing: Dull razor method \cite{dull}}
\item{Segmentation: Conventional Canny, B-Otsu and Chan-Vese methods and the proposed Parameter Selection  Method (PSM) and Maximum Area Method (MAM)}
\item{Postprocessing: Morphological closing, Hole filling}
\item{Feature extraction: ABC criteria}
\item{Classification:  Weighted ROC thresholding}
\end{itemize}
The segmentation step will be explained in detail in the next section, with a particular concentration on the suggested two segmentation methods: PSM and MAM.  The other two major steps will also be explained briefly. 

Among the many segmentation methods used for diagnosing skin cancer,  we considered the following methods: the Canny method \cite{canny}, (B-channel) Otsu method \cite{otsu, otsu1,otsu2}, and Chan-Vese method \cite{ChanVese,VeseChan, active2, active1, Adjed}. These represent the most widely used techniques among respective difference, histogram, and region based sementation methods.  In addition, we propose PSM and MAM in this paper. PSM segments a lesion by selecting appropriate parameters for high-boost filtered images. The maximum size segmented image is then selected in MAM among those analyzed by PSM, PSMW(Parameter Selection Method after Whole image normalization), PSMB(Parameter Selection Method after background normalization).  

PSM is based on the B-Otsu method. The blue channel is known to be the best color channel used in segmentation \cite{blue}. However, lesions segmentated by the B-Otsu method generally turn out smaller than those by a human expert's segmentation \cite{TMI}  or by many other conventional segmentation methods \cite{Overview4,SegmentationHybrid}. To overcome this disadvantage, the high-boost filter is used to maximize a lesion's inhomogeneity compared to the background homogeneity of the blue channel. In this way, PSM is used to select the best parameter for the high-boost filters. 

The 900 samples used in this paper  \cite{ISIC} are too diverse, as such, some normalizations are required. These are whole (blue-channel) image normalization and  background image normalization. PSM  is applied to choose the best parameters after two whole and background normalizations are carried out. These normalizations are called PSMW and PSMB, respectively. MAM is used to select the maximum size image among segmentation images from PSM, PSMW  (PSM after Whole image normalization), and PSMB  (PSM after Background image normalization). This maximum size selection is also based on a report that shows conventional segmentation images are generally smaller than human experts' segmentation images \cite{Overview4,SegmentationHybrid} and also on our observations that PSM segmentation images are smaller than the expert's segmentation shown in \cite{ISIC}.


We extracted features from the seperated lesion to differentiate malignant from benign melanomas. We used  the ABCD criteria \cite{abcdcriteria} rather than Menzis \cite{Menzis} or 7-point checklist \cite{7p2} methods in the Feature Extraction step. The ABCD criteria extracts the following four features: Asymmetry of the lesion shape, Border irregularity of the lesion, Color variegation of the non-uniform lesion, and Diameter of the lesion being greater than 6mm.  In this paper, mathematically quantified values  \cite{TMI}, which improve on the results of \cite{mine},  will be used for the ABCD criteria.

Many classification methods diagnosing malignant melanoma have been studied such as  total dermoscopy score (TDS) \cite{Kel}, k-nearest neighbor (kNN) \cite{knnann}, support vector machine (SVM) \cite{SVM1}, artificial neural network (ANN) \cite{ann}, and median thresholding method \cite{mine}. We will use the weighted ROC thresholding methods \cite{TMI} to select the threshold values by considering the ROC (Receiver Operating Characteristic) space with respect to the weighted sum of the features A,B, and C in the ABCD criteria. In this method, the best threshold value along with the weights are selected in such a way that the  sensitivity was highest among the the threshold values whose sensitivity and specificity were closest to (1,1) in the ROC space for every weighting. Note that sensitivity is more important than specificity with respect to the risk of malignant melanoma. 

\section{Materials and Methods}
The three conventional and two new, proposed segmentation methods and their influences on melanoma diagnosis will be explored in this section.

For evaluation of the segmentation alone without any following classification procedure,  we used Jaccard indices to evaluate the various segmentation methods while using the skin cancer expert's segmentation \cite{ISIC} as the ground truth. To  compare the efficiency of any two segmentation methods, we use the following notation: 
\begin{itemize}
\item{$J_1$ : Jaccard index for `Method 1',}
\item{$J_2$ : Jaccard index for `Method 2',}
\item{${\displaystyle J_{1,2} = \dfrac{J_1-J_2}{\max(J_1,J_2)}}$: Jaccard index comparison parameter (of `Method 1'  against `Method 2') .}
\end{itemize}
Given an assigned value $\delta$, we say that
\begin{itemize}
\item{`Method 1' is better than `Method 2'   if $J_{1,2}>\delta$,}
\item{`Method 2' is better than `Method 1'   if $J_{1,2}<-\delta$,}
\item{`Method 1' is similar to `Method 2'   if $|J_{1,2}|\le \delta$.}
\end{itemize}
Furthermore, we also used the accuracy, sensitivity, and specificity of the melanoma classification to compare with the two suggested, three conventional, and  the expert's segmentations.  

\subsection{Chan-Vese method (CV) \cite{ChanVese}}
 This is a well-known region based segmentation method and is based on the theories of active contours, curve evolution, the Mumford-Shah functional, and level sets. This method has been applied to skin cancer detection in \cite{Adjed}. A multiphase model by Vese and Chan has also been used in \cite{VeseChan}. We used the algorithm as given in \cite{CVMatlab}.
 
\subsection{Canny method \cite{canny} }
The Canny method is a well-known difference based method. This method is used in \cite{mine} and named  the `Edge-imfill method' in the paper.

\subsection{B-Otsu method}
We applied the Otsu method \cite{otsu}, which is a well-known one-dimensional histogram based method,  to the B-channel of the RGB color system. The B-channel is known to be the best channel with which  to segment the lesion of interest \cite{blue}. In \cite{TMI}, we applied the Otsu method to the U channel of the YUV color system, which is similar to the B channel in the RGB system. 

\subsection{Parameter Selection Method (PSM) }
\begin{figure}[H]
\includegraphics[width=16cm,height=10cm]{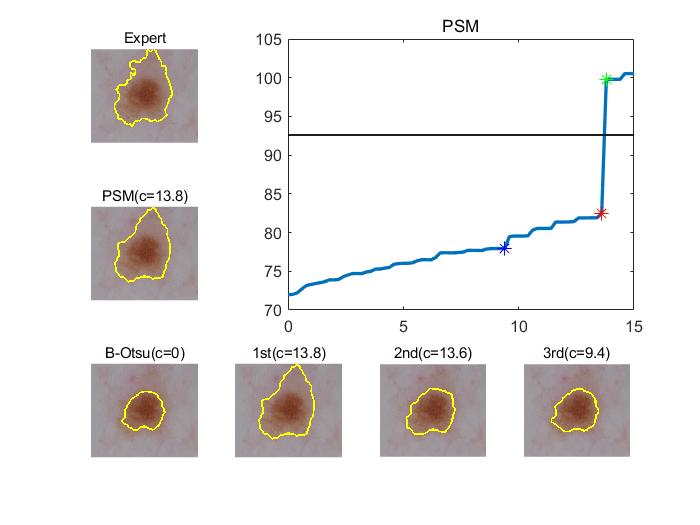}
\caption{The process of obtaining PSM segmented image.} 
\label{fig:PSM}
\end{figure}

The B-Otsu method tends to give smaller segmentation lesions compared to human experts' segmentations, which is typical of conventional segmentation methods
\cite{Overview4,SegmentationHybrid}. To improve the lesion size given by the B-Otsu method, we added some perturbation to segmented lesions, which are relatively nonuniform compared to the background.  The perturbation is added by using high-boost filter as follows:  
$$
H = Id + c*Laplacian = \left(
\begin{array}{ccc}
0 & -c & 0 \\
-c & 4c+1 & -c\\
0 & -c & 0
\end{array}
\right).
$$
We suggest the following strategy to select the best parameter $c$ in the high-boost filter $H$.
\begin{enumerate}
\item{FOR c=0:$\Delta c$:$c_{max}$\\
 Apply high-boost filter to the B-channel, segment the filtered image using Otsu method, and 
calculate the mean of the segmented lesion. Let us denote the mean in terms of the parameter $c$ by $M(c)$.\\
END}
\item{Select three values $c_1,c_2,c_3$ where $M^{''}(c)$ is highest. }
\item{Calculate five means of lesions segmented by the B-Otsu method for the high-boost  filtered images with parameters $0,c_1,c_2,c_3$ and the k-means clustering method with $k=2$. Select one parameter from $0,c_1,c_2,c_3$, the mean at which is closest to the mean by the 2-means method.}
\end{enumerate}
Note that $M^{''}$ captures the change of the mean value increase rate, rather than the increase itself.  Since the lesion segmented by the B-Otsu method or most other conventional methods is generally smaller than dematologists' segmentation \cite{Overview4,SegmentationHybrid}, the segmented lesion is required to be enlarged onto the background. Due to the perturbation of the lesion by the high-boost filter,  the segmented lesion is generally enlarged with respect  to parameter $c$, although there can be some exceptional contraction. To maximize the enlargement, we select three values as in Fig. \ref{fig:PSM}, the mean increase rate at which is most rapid. However, it is possible that the enlargement could come from some noise due to the perturbation and, therefore, the segmentation lesion from the three values and $c=0$ might deviate from the ground truth. Therefore, we select  a 2-means segmentation image as the ground truth to measure how much the high-boost image has deviated by comparing the means of the B-channel.  Note that a 2-means segmentation images are not adequate to be used in the segmentation of melanoma lesion since usually these are not connected, but the mean of the B-channel image could be a good standard from which to obtain a connected segmentation image.  

\subsection{Maximum Area Method(MAM)}
\begin{figure}[H]
\centering
\includegraphics[width=8cm,height=2cm]{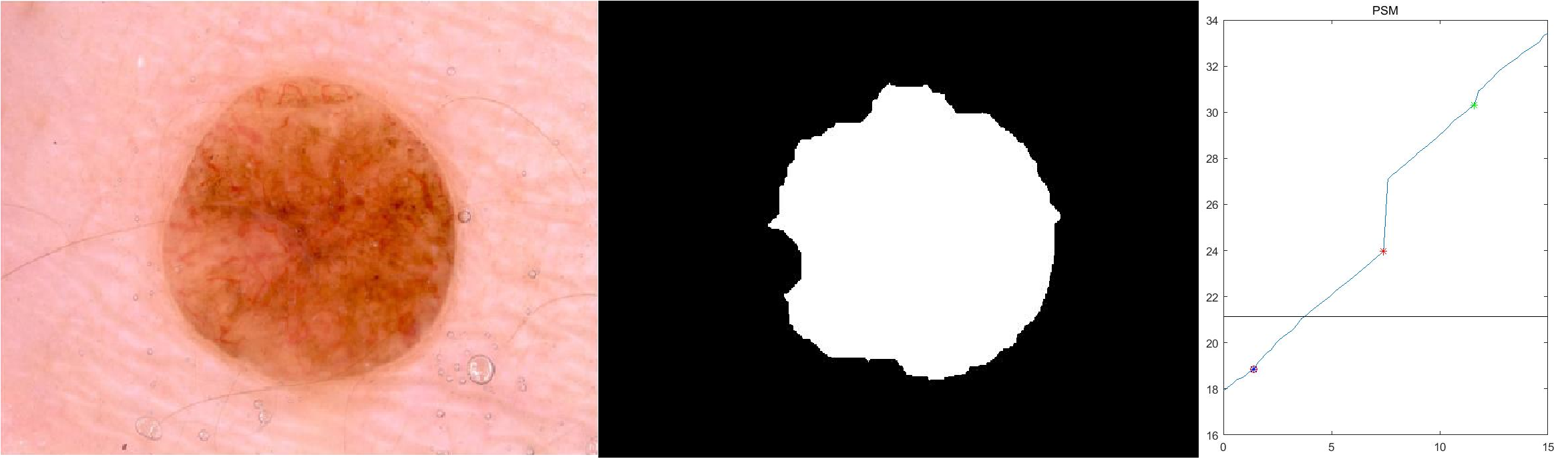}
\\\centering\text{(a)}\\
\includegraphics[width=8cm,height=2cm]{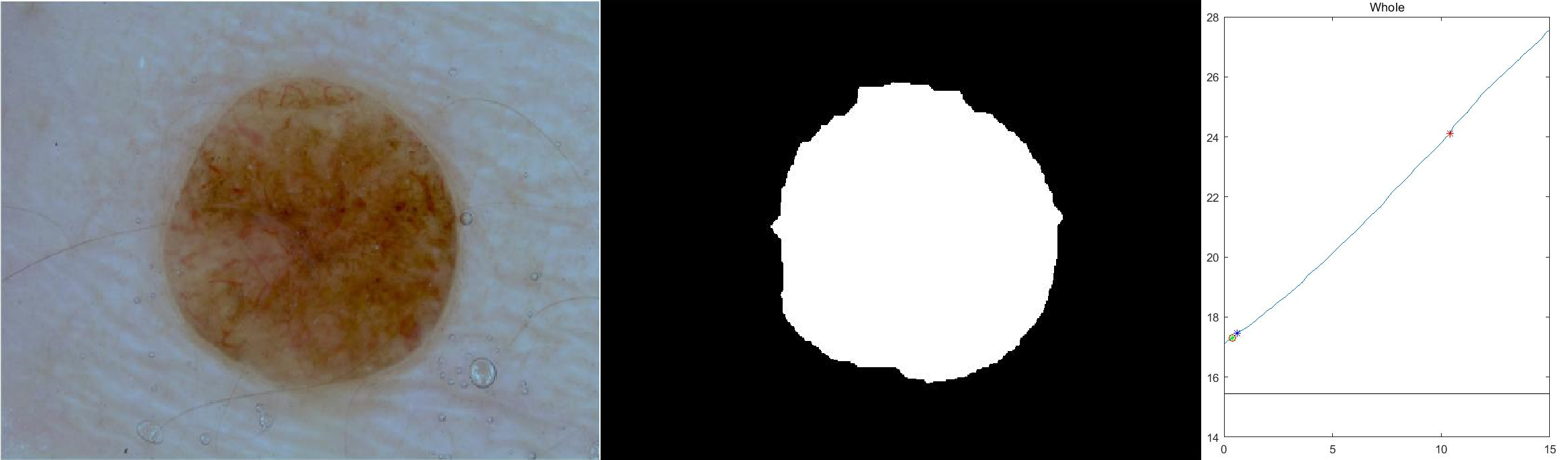}
\\\centering\text{(b)}\\
\includegraphics[width=8cm,height=2cm]{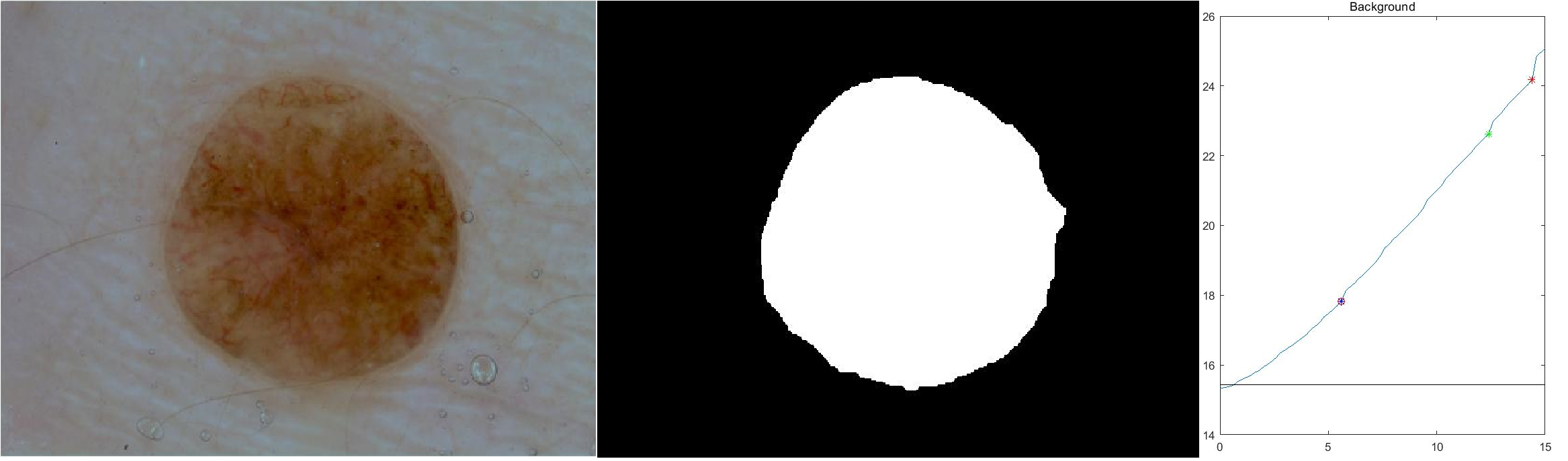}
\\\centering\text{(c)}\\
\caption{The process of obtaining MAM segmented image: (a)PSM (b)PSMW (c) PSMB. In this case, PSMB is chosen as it has the maximum area.} 
\label{fig:MAM}
\end{figure}

Since the 900 samples that we have taken from \cite{ISIC} to test our system are diverse, some kind of normalization is required. We used two kinds of normalization:  whole image normalization and background normalization. This is followed by selecting the $c$ value as in PSM.  Therefore, we get three segmented images by PSM, PSMW, and PSMB. Even though most segmented images by PSM are larger than those from B-Otsu, they are still smaller than the expert's segmentation. Therefore, we use Maximum Area Method to select the maximum segmented lesion among those generated by PSM, PSMW, and PSMB, as shown in Fig. \ref{fig:MAM}.  The followings are the details of the algorithm for MAM given some real number $\epsilon$:

\begin{enumerate}
\item{PSM: Implement PSM segmentation.}
\item{Let the mean of the whole image and background image in the B channel be $m_W$ and $m_B$, respectively.}
\item{ PSMW: Multiply the B- channel by $2^{n-1}(1+\epsilon)/m_W$. Here, $n$ is the number of bits representing the B-channel. Implement the same procedure as in PSM for the multiplied image.}
\item{PSMB: Multiply the B- channel by $2^{n-1}(1+\epsilon)/m_B$. Implement the same procedure as in PSM for the multiplied image.}
\item{Select the image that has the maximum area among PSM, PSMW, and PSMB segmented images.} 
\end{enumerate}

PSMW and PSMB have the following advantage over PSM. Values larger than $2m_W/(1+\epsilon)$ and $2m_B/(1+\epsilon)$ for PSMW and PSMB, respectively, are saturated into $2^n$. Due to this saturation, noises with higher values (such as salt in the salt-and-pepper noise) have been made uniform.

\section{Results}                                                                                                                                                                                                                                                                                                                                                                                                                                                                                                                                                                                                                                                                                                                                                                                                                                                                                                                                                                                                                                                                                                                                                                                                                                                                                                                                                                                                                                                                                                                                                                                                                                                                                                                                                                                                                                                                                                                                                                                                                                                                                                                                                                                                                                                                                                                                                                                                                                                                                                                                                                                                                                                                                                                                                                                                                                                                                                                                                                                      
\begin{figure}[H]
\includegraphics[width=8cm,height=6cm]{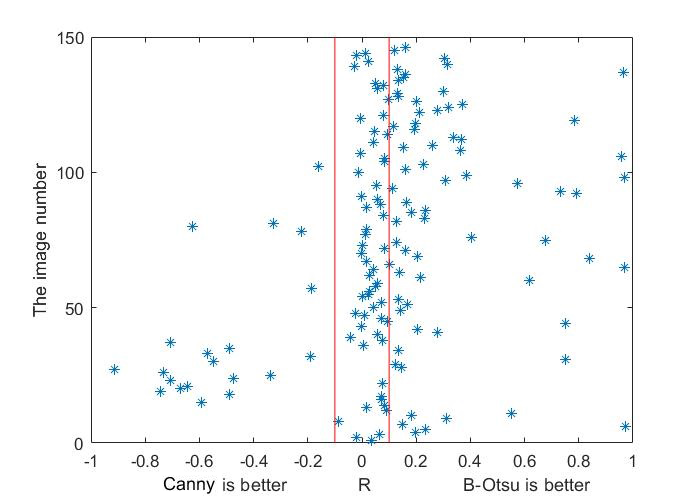}
\includegraphics[width=8cm,height=6cm]{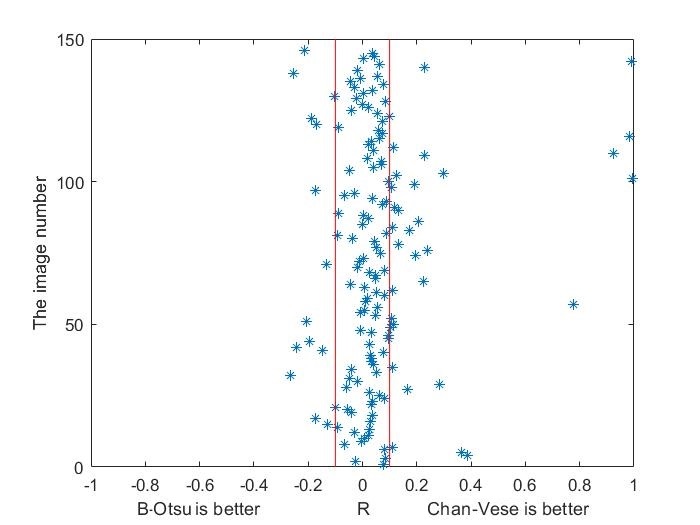}
\centering\text{(a)\qquad\qquad\qquad\qquad\qquad(b)}\\
\includegraphics[width=8cm,height=6cm]{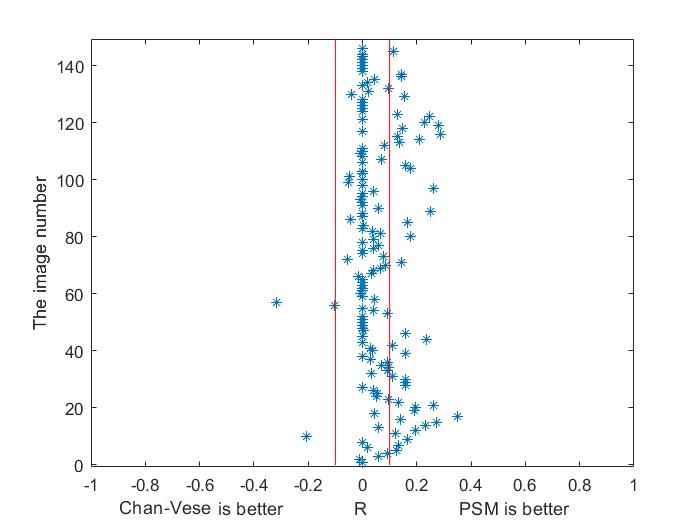}
\includegraphics[width=8cm,height=6cm]{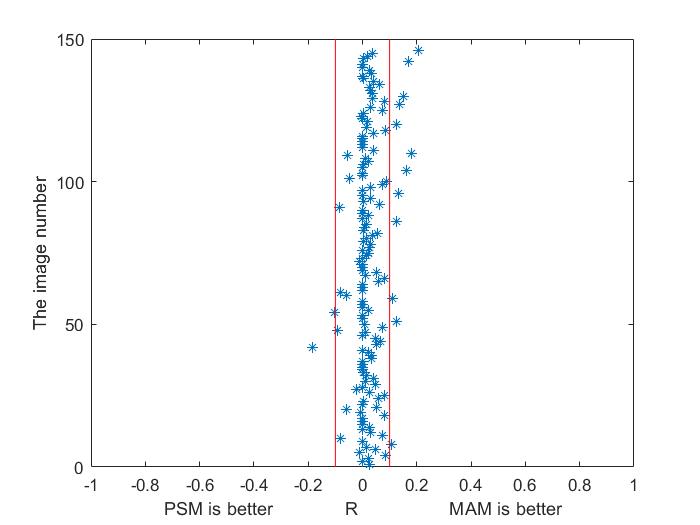}
\centering\text{(c)\qquad\qquad\qquad\qquad\qquad(d)}
\caption{The Jaccard index comparison parameters between (a) Canny and B-Otsu method, (b) B-Otsu and Chan-Vese method, (c) Chan-Vese method and PSM, and (d) PSM and MAM for the 146 images.  The red line represents $\delta=0.1$ and $-0.1$.} 
\label{fig:synchro}
\end{figure}

Three conventional (Canny, Chan-Vese, B-Otsu) methods and two proposed methods (PSM and MAM), are compared using 900 (724 benign and 176 malignant) skin samples taken from \cite{ISIC}. Since the samples have different sizes, we contracted the samples into images with less than 12,000 pixels.  In Fig. \ref{fig:synchro} and Table \ref{tab:synchro}, 146 (75 benign and  71 malignant) from the  900 samples were chosen for simple computation.

Matlab 19a is used to implement the five methods.  We used 1000 iterations and the and initial condition `Large'  for the Chan-Vese method, $c_{max}=15, \Delta c =0.2$ for PSM, $\epsilon=0$ for MAM, and $\delta=0.1$ for the Jaccard index comparison parameter.                                                                                                                                                                                                                                                                                                                                                                                                                                                                                                                                                                                                                                                                                                                                                                                                                                                                                                                                                                                                                                                                                                                                                                                                                                                                                                                                                                                                                                                                                                                                                                                                                                                                                                                                                                                                                                                                                                                                                                                                                                                                                                                                                                                                                                                                                                                                                                                                                                                                                                                                                                                                                                                                                                                                                                                                                                                                                                                                                                                                                                                                                                                                                                                                                                                                                                                                                                                                                                                                                                                                                                                                                                                                                                                                                                                                                                                                                                                                                                                                                                                                                                                                                                                                                                                                                                                                                                                                                                                                                                                                                                                                                                                                                                                                                                                                                                                                                                                                                                                                                                                                                                                                              

Over the 146 samples, we found that MAM is better than PSM, PSM is better than Chan-Vese, Chan-Vese is better than B-Otsu, and B-Otsu is better than Canny method in terms of the Jaccard index comparison parameter, as shown in Fig. \ref{fig:synchro} and Table \ref{tab:synchro}.

\begin{figure}[H]
\includegraphics[width=12cm,height=4.5cm]{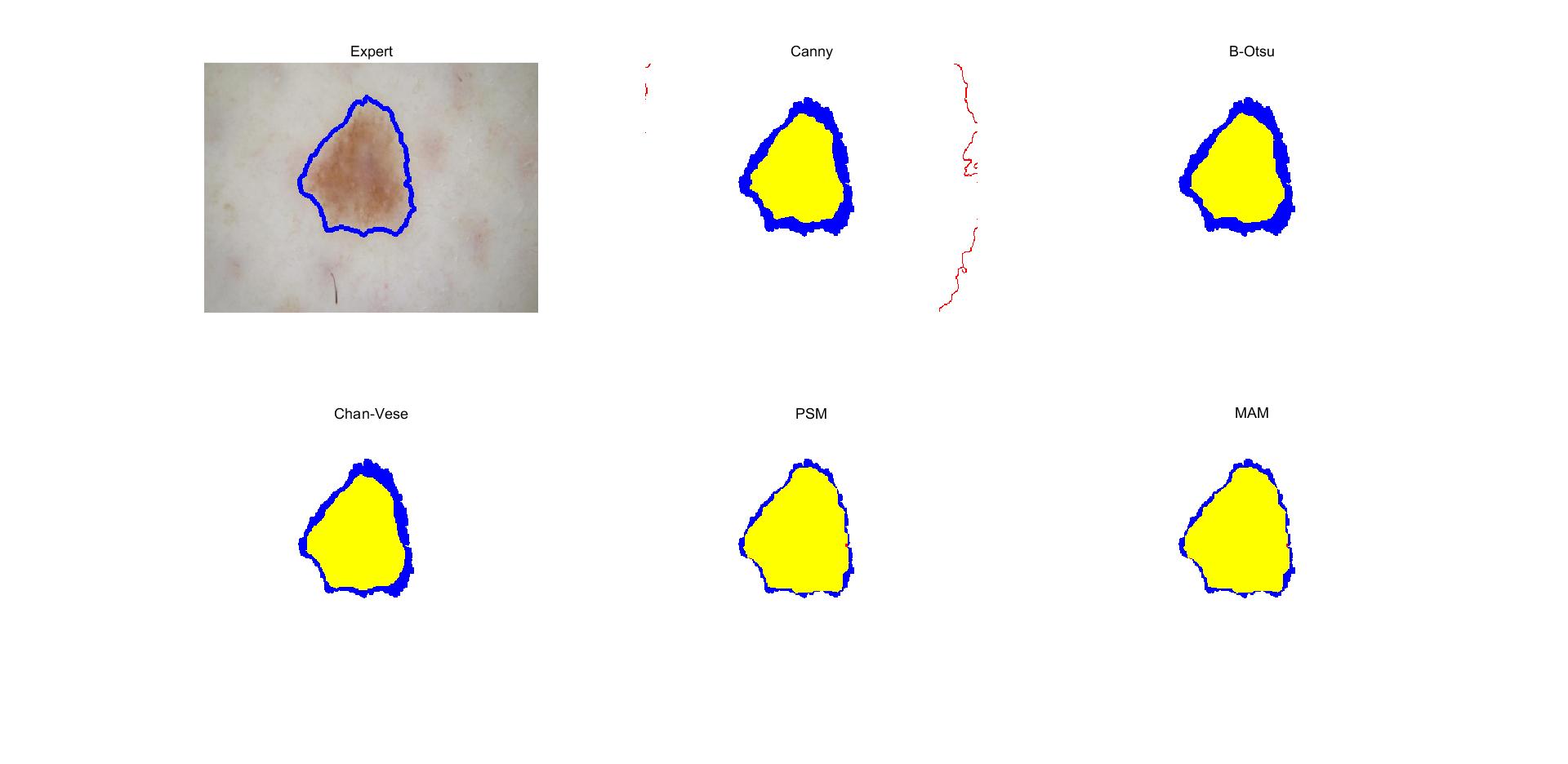}\\
\centering\text{(a)}\\
\includegraphics[width=12cm,height=4.5cm]{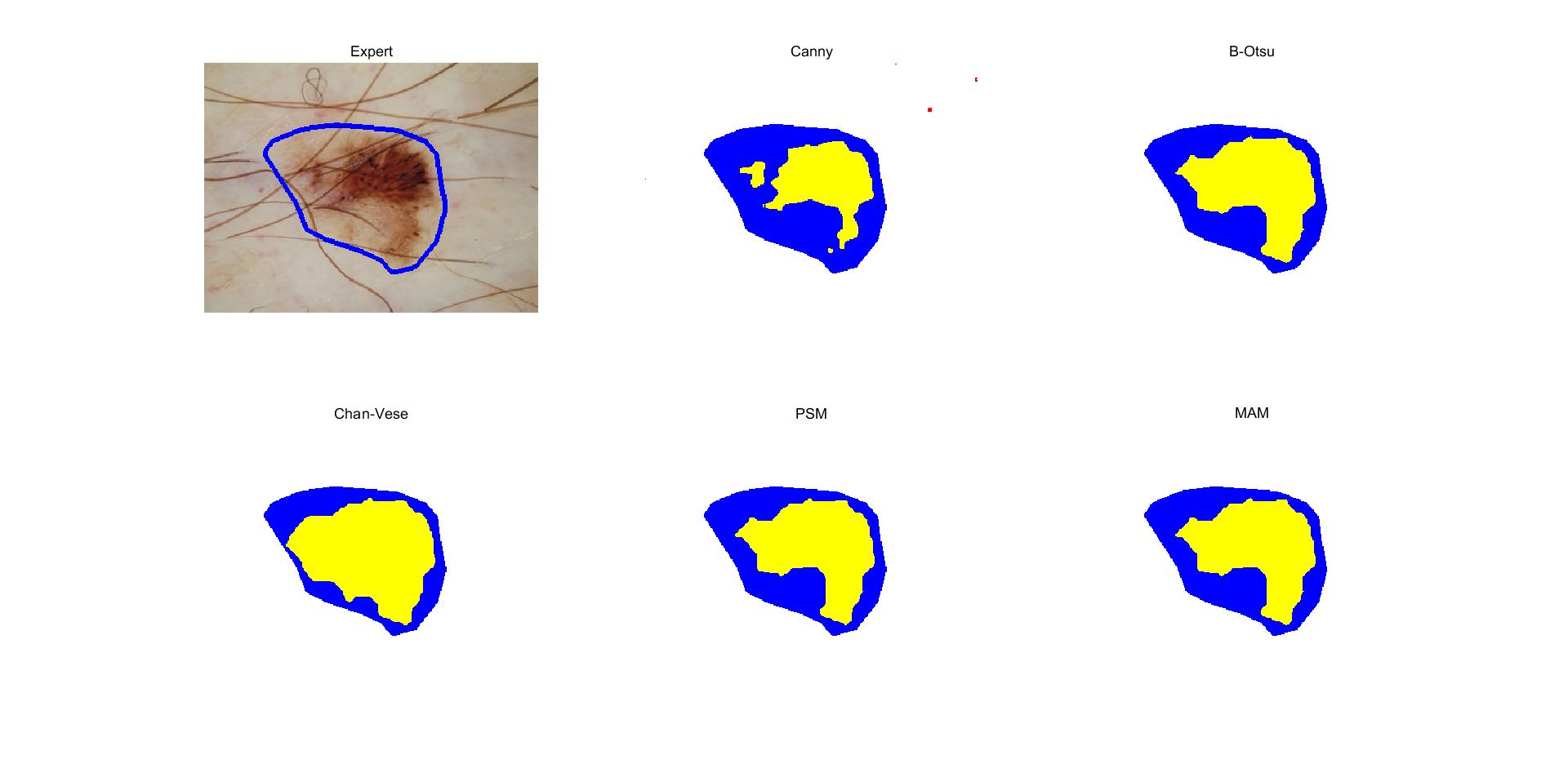}\\
\centering\text{(b)}\\
\includegraphics[width=12cm,height=4.5cm]{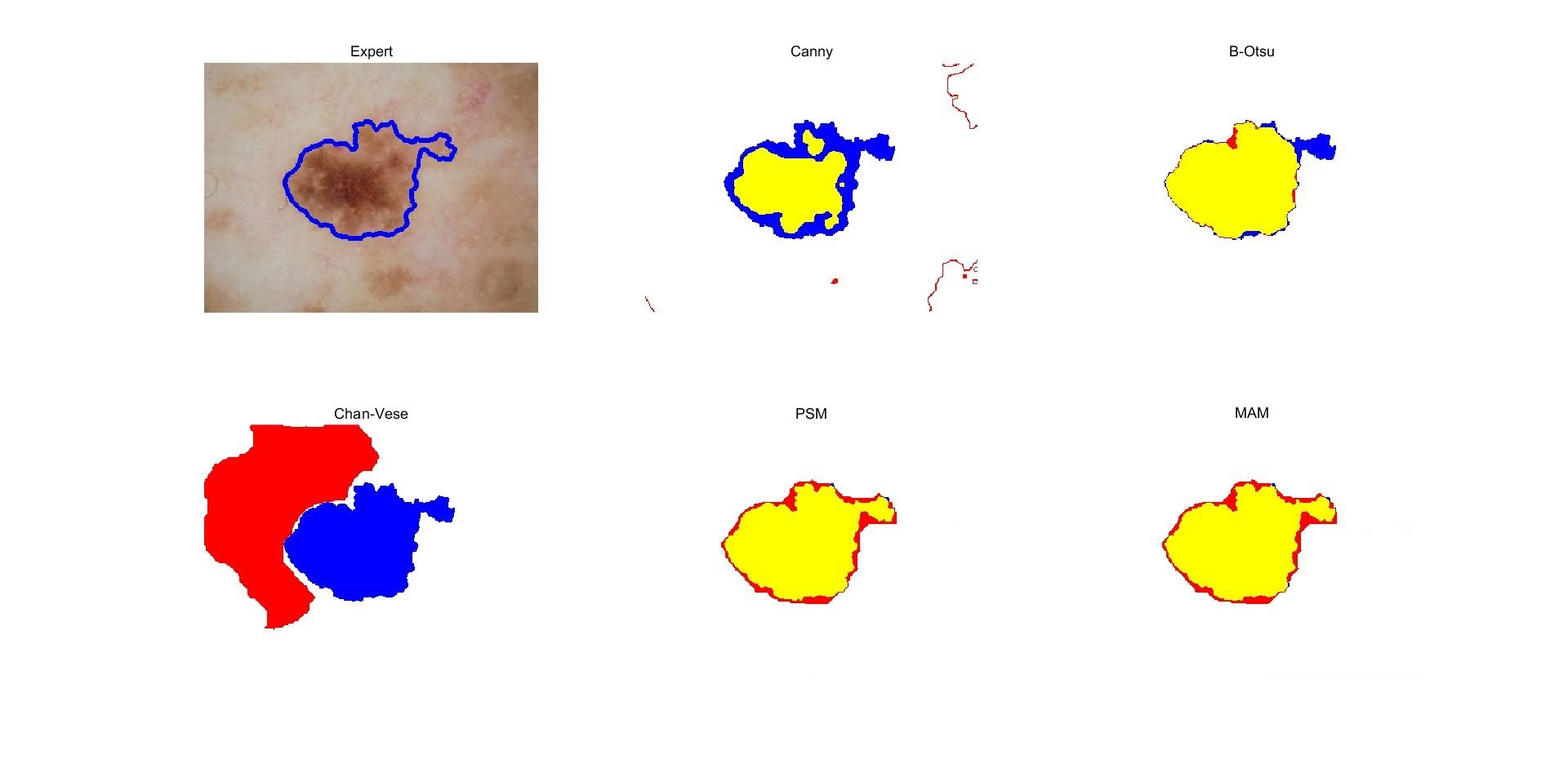}\\
\centering\text{(c)}\\
\includegraphics[width=12cm,height=4.5cm]{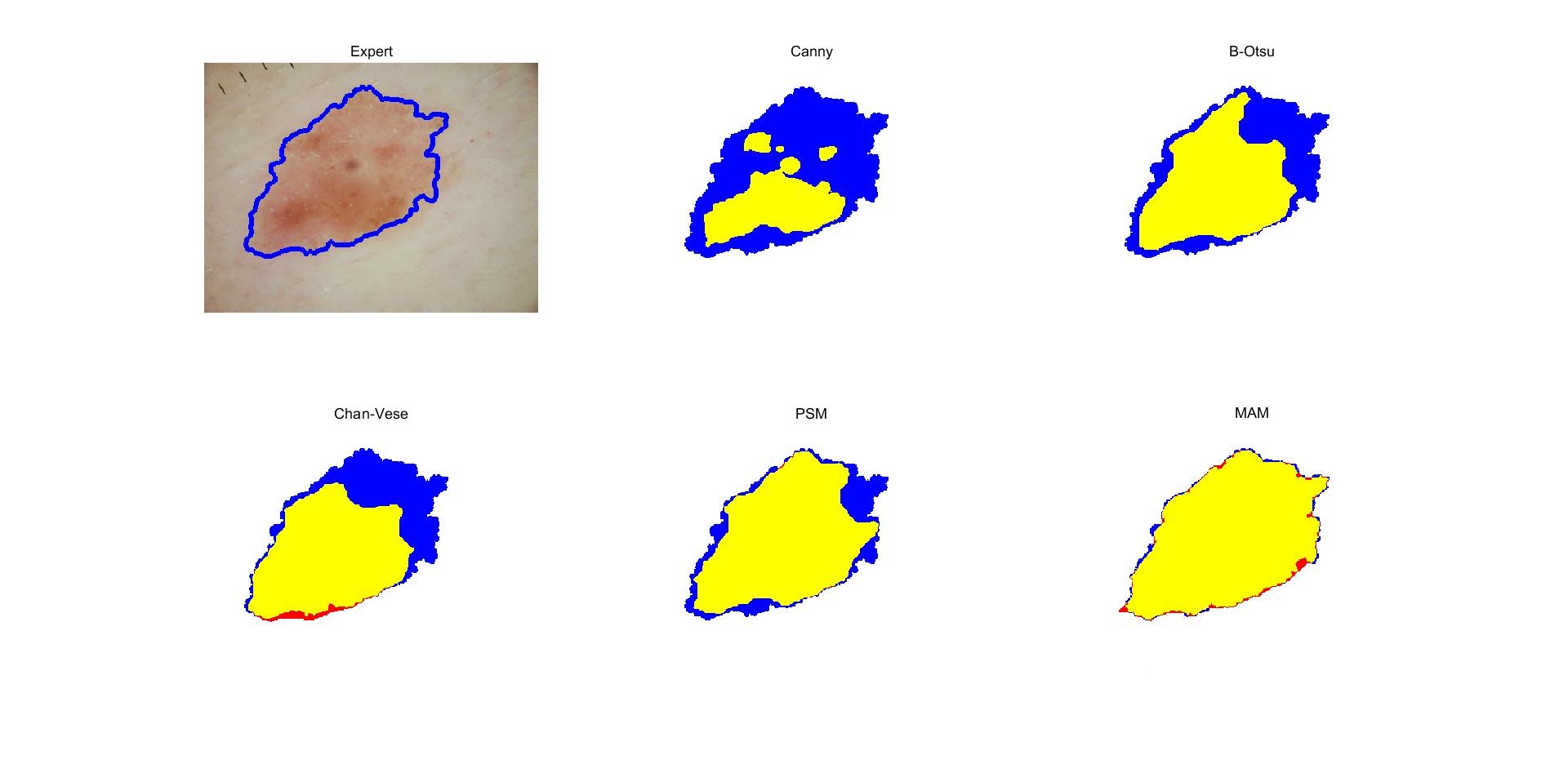}\\
\centering\text{(d)}
\caption{Comparisons of the five segmentation methods with respect to the expert's segmentation as the ground truth.} 
\label{fig:comp}
\end{figure}

\begin{table}[H]
\begin{centering}
\begin{tabular}{|c|ccc|c|}
\hline
{}& better & similar & worse & {}\\
{}& than           & to           & than &{}\\
\hline 
B-Otsu is& 69 &57 & 20 & Canny\\
\hline
Chan-Vese  is&31 &101& 14& B-Otsu\\ 
\hline
PSM is& 26& 117& 3& Chan-Vese\\
\hline
MAM is& 12& 132& 2& PSM\\
\hline
\end{tabular}
\caption{Jaccard index comparison for the five segmentation methods in Fig. \ref{fig:synchro}. }
\label{tab:synchro}
\end{centering}
\end{table} 
                                                                                                                                                                                                                                                                                                                                                                                                                                                                                                                                                                                                                                                                                                                                                                                                                                                                                                                                                                                                                                                                                                                                                                                                                                                                                                                                                                                                                                                                                                                                                                                                                                                                                                                                                                                                                                                                                                                                                                                                                                                                                                                                                                                                                                                                                                                                                                                                                                                                                                                                                                                                                                                                                                                                                                                                                                                                                                                                                                                                                                                                                                                                                                                                                                                                                                                                                                                                                                                                                                                                                                                                                                                                                                                                                                                                                                                                                                                                                                                                                                                                                                                                                                                                                                                                                                                                                                                                                                                                                                                                                                                                                                                                                                                                                                                                                                                                                                                                                                                                                                                                                                                                                                                                                                                                                                                                                                                                                                                                                                                                                                                                                                                                                                                                                                                                                                                                                                                                                                                          In Fig. \ref{fig:comp}, we show some examples of the five segmentation methods with respect to the expert's segmentation as the ground truth. Let the lesion segmented by one of the five ethods and by the expert be $S$ and $E$, respectively. The color in the figure represent the followings:
\begin{itemize}
\item{Yellow: $S\cap E$}
\item{Blue: $E\setminus S$}
\item{Red: $S\setminus E$}
\end{itemize}                                                                                                                                                                                                                                                                                                                                                                                                                                                                                                                                                                                                                                                                                                                                                                                                                                                                                                                                                                                                                                                                                                                                                                                                                                                                                                                                                                                                                                                                                                                                                                                                                                                                                                                                                                                                                                                                                                                                                                                                                                                                                                                                                                                                                                                                                                                                                                                                                                                                                                                                                                                                                                                                                                                                                                                                                                                                                                                                                                                                                                                                                                                                                                                                                                                                                                                                                                                                                                                                                                                                                                                                                                                                                                                                                                                                                                                                                                                                                                                                                                                                                                                                                                                                                                                   The segmentation by the Canny method might have unexpected curves which the other methods do not, as seen in Fig.\ref{fig:comp} (a),(b),(c) and has a much smaller image than the expert's, as seen in Fig.\ref{fig:comp} (a),(b),(c),(d). Overall, the five segmentation images are smaller than the expert's and become bigger in the order: Canny, B-Otsu, Chan-Vese, PSM, MAM. The segmentation images from Chan-Vese are better than PSM or MAM images in some cases, as shown in  Fig. \ref{fig:comp}(b), but  does not match the expert's segmentation in other cases, as shown in Fig. \ref{fig:comp} (c), since it is a region based segmentation method that is highly dependent on the initial guess and iteration number. 

The Jaccard index of the 900 samples for the five methods using an expert's segmentation as the ground truth show  the methods ordered from best to worst are: MAM, PSM, Chan-Vese, B-Otsu, and Canny, as shown in Table \ref{tab:seg900}.  We also tested classification efficiency of the five methods using the 900 samples along with the expert's segmentation followed by the suggested feature extraction and thresholding steps in Section 1. The accuracy, specificity, and sensitivity of the five methods shows a similar tendency as the Jacard index result with two exceptions: The accuracy and specificity of PSM were a little better than MAM.  

\begin{figure}[H]
\includegraphics[width=8.5cm,height=1.5cm]{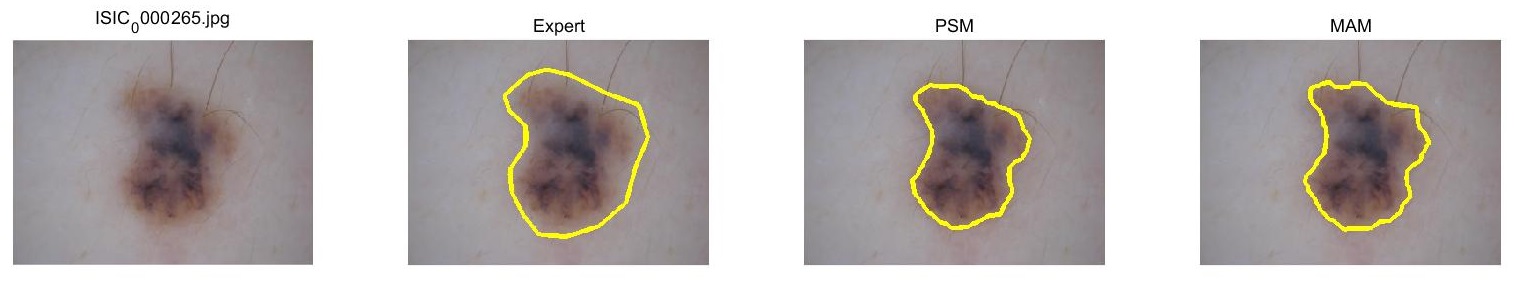}\\
\centering\text{(a)}\\
\includegraphics[width=8.5cm,height=1.5cm]{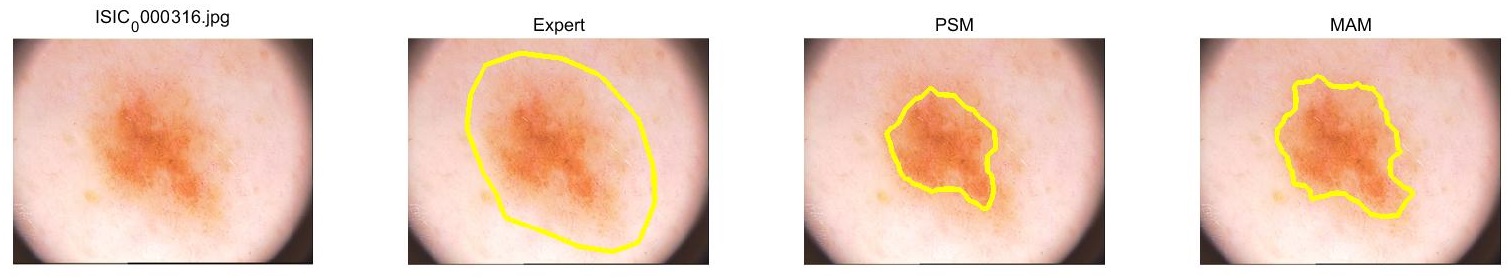}\\
\centering\text{(b)}\\
\includegraphics[width=8.5cm,height=1.5cm]{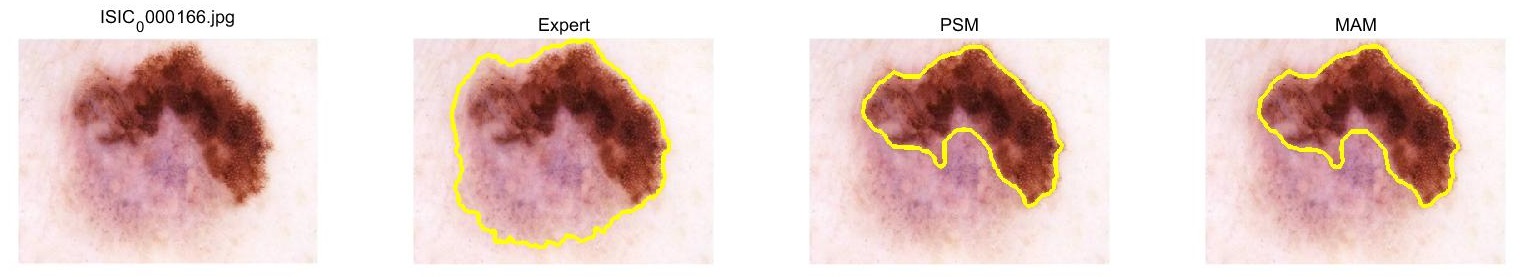}\\
\centering\text{(c)}\\
\includegraphics[width=8.5cm,height=1.5cm]{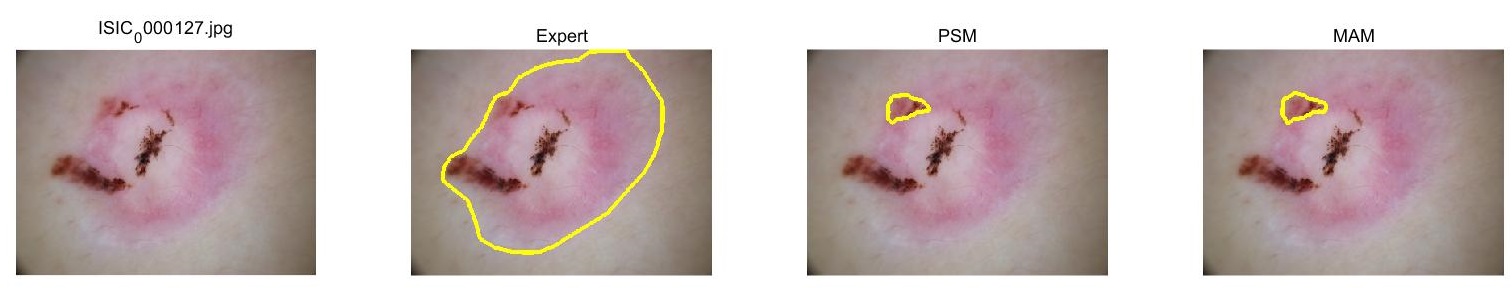}\\
\centering\text{(d)}\\
\includegraphics[width=8.5cm,height=1.5cm]{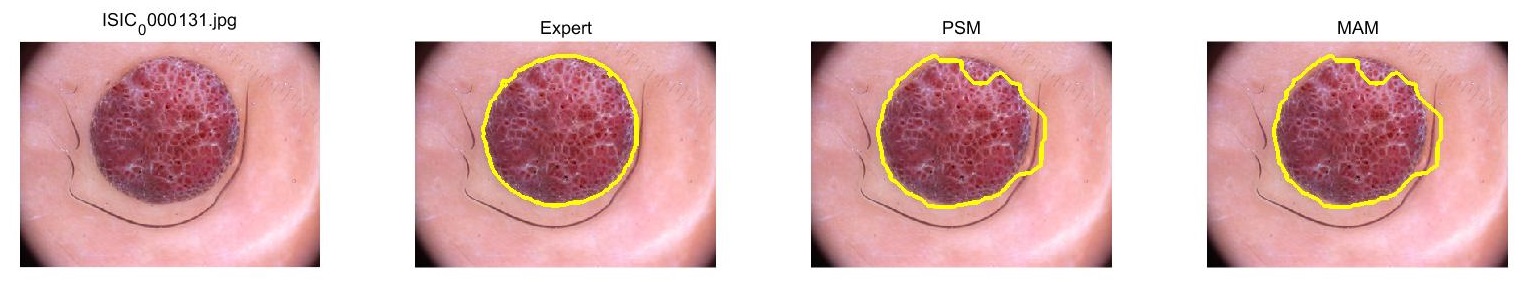}\\
\centering\text{(e)}\\
\includegraphics[width=8.5cm,height=1.5cm]{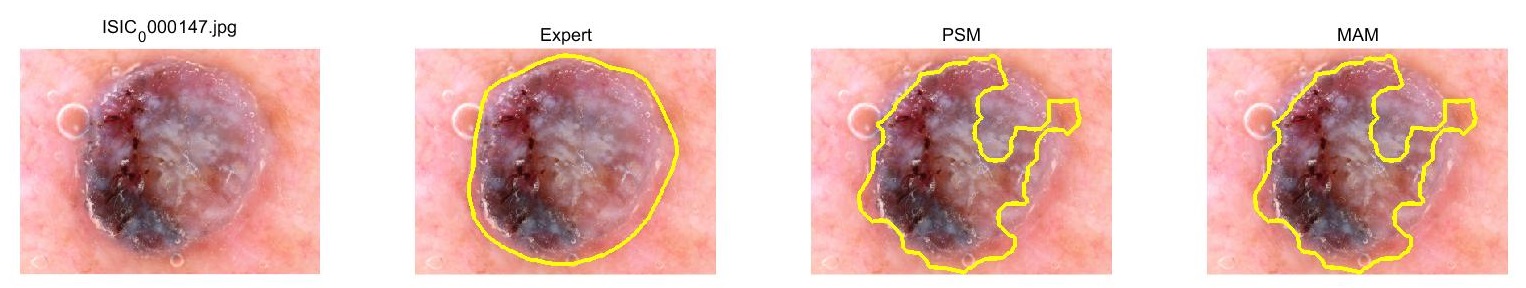}\\
\centering\text{(f)}\\
\caption{Comparison of segmentation results with the expert, PSM, and MAM.} 
\label{fig:Worse}
\end{figure}

\begin{figure}[H]
\includegraphics[width=1.5cm,height=1.5cm]{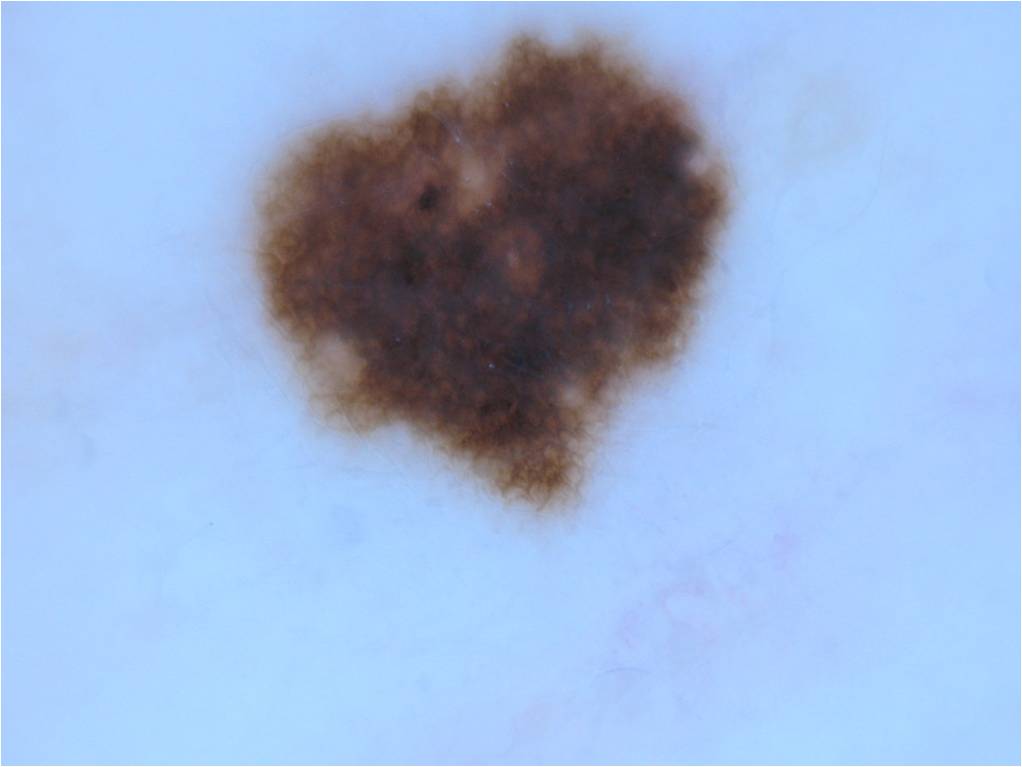}
\includegraphics[width=1.5cm,height=1.5cm]{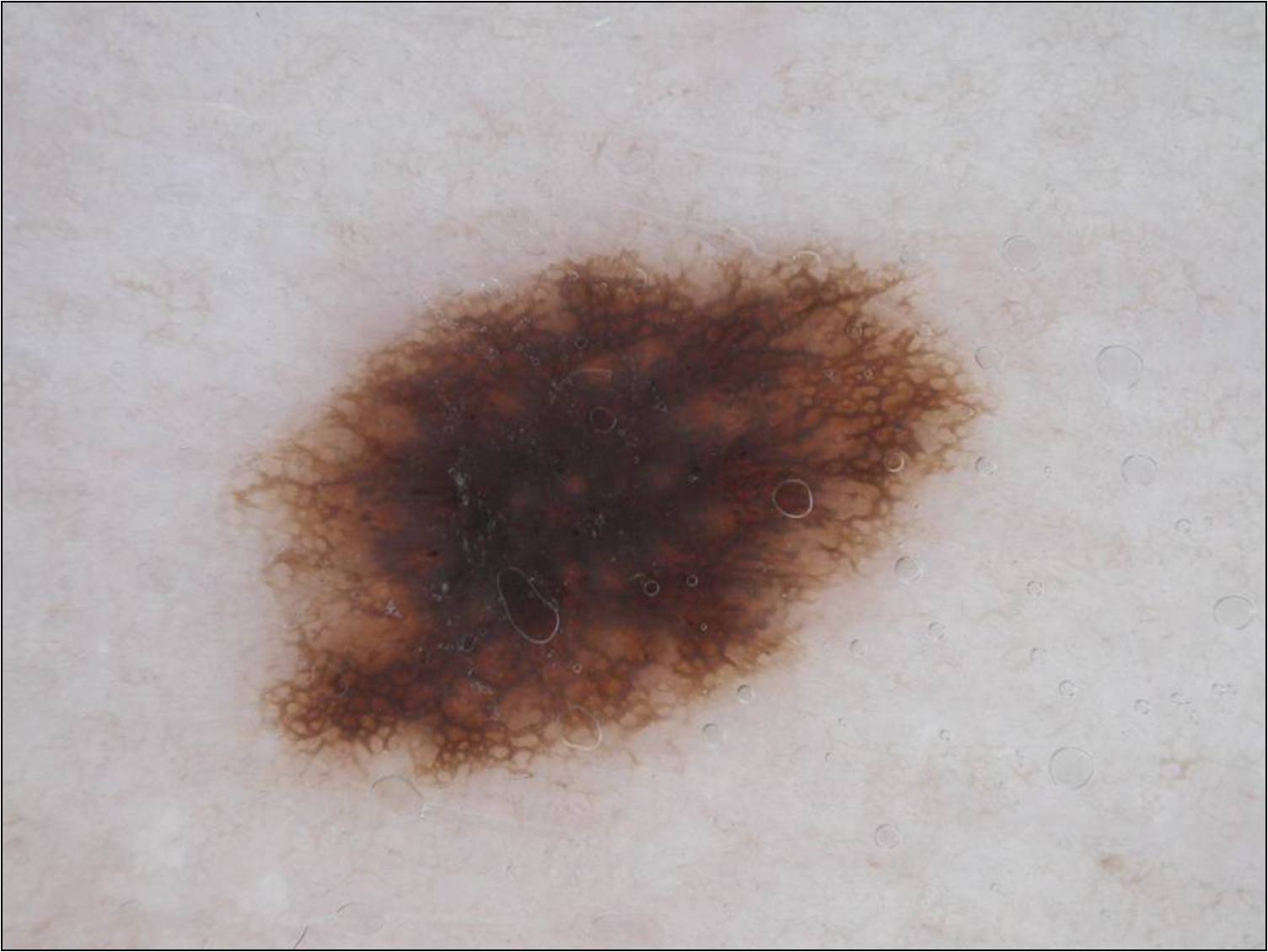}
\includegraphics[width=1.5cm,height=1.5cm]{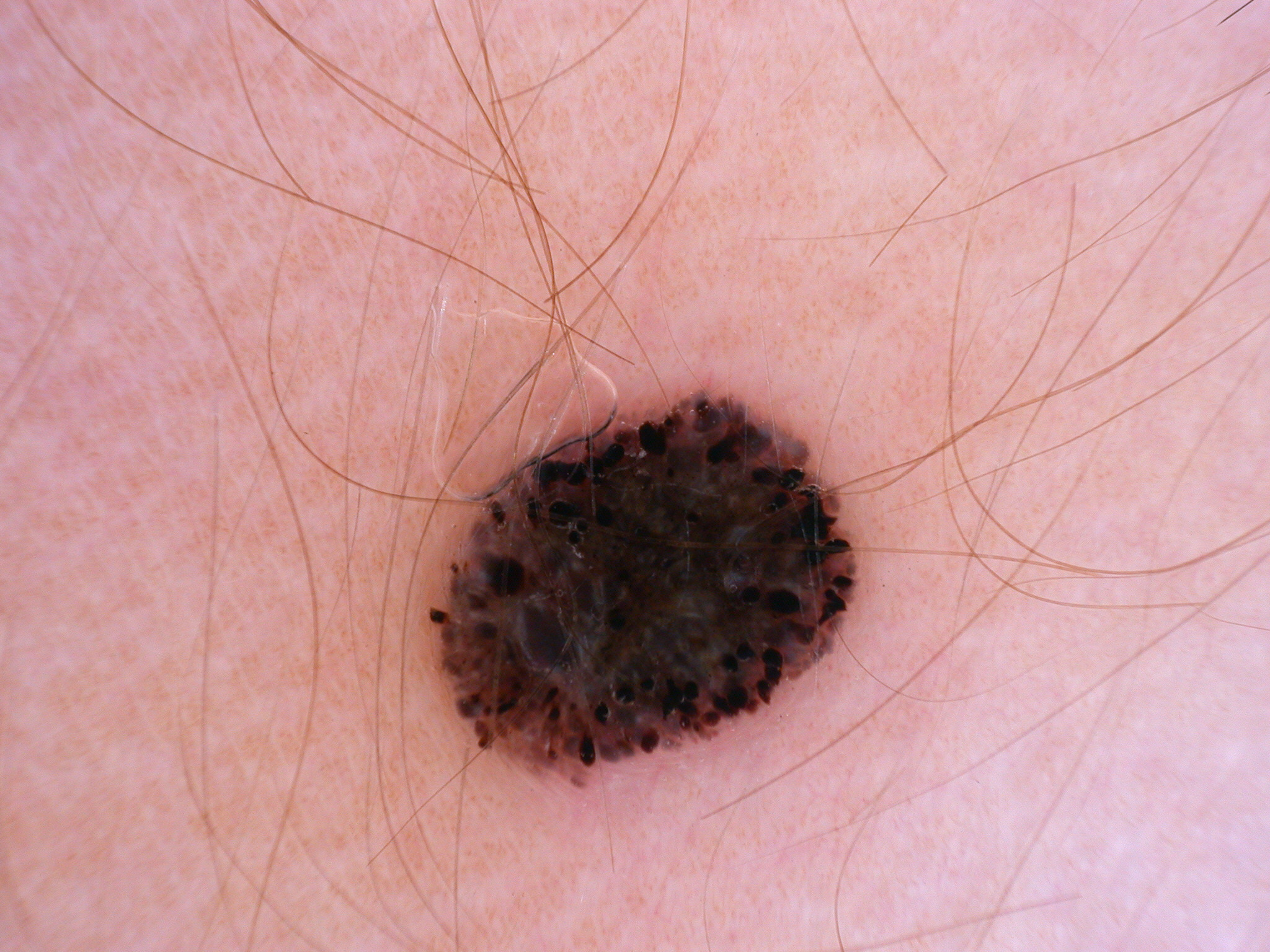}
\includegraphics[width=1.5cm,height=1.5cm]{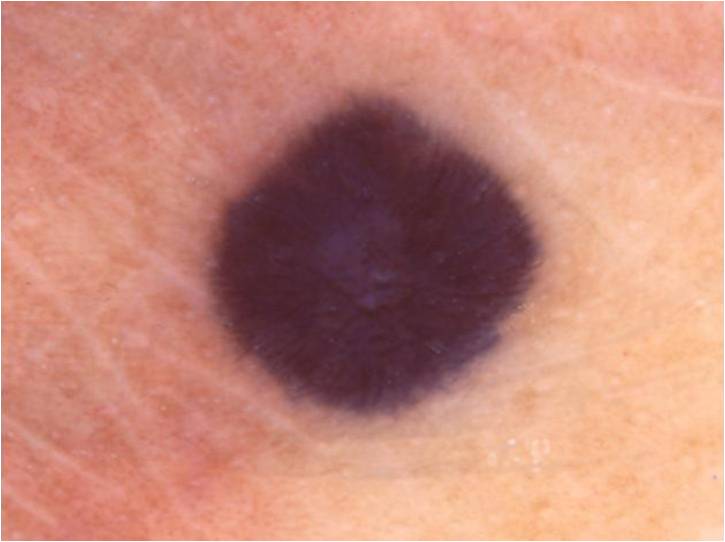}
\includegraphics[width=1.5cm,height=1.5cm]{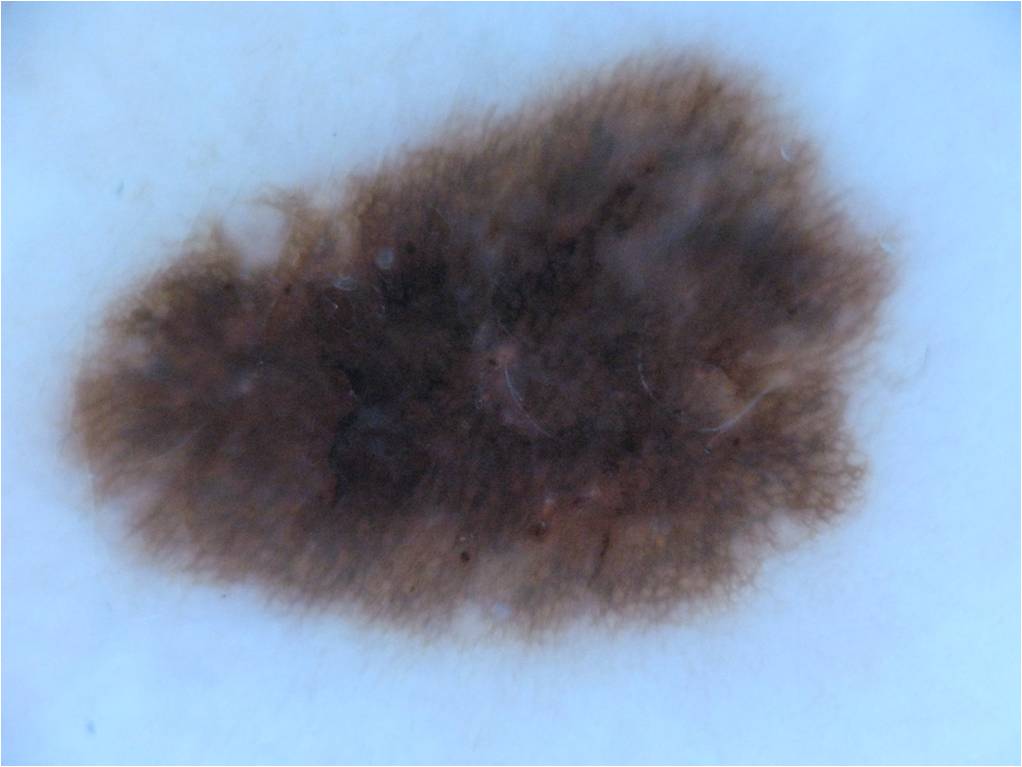}
\includegraphics[width=1.5cm,height=1.5cm]{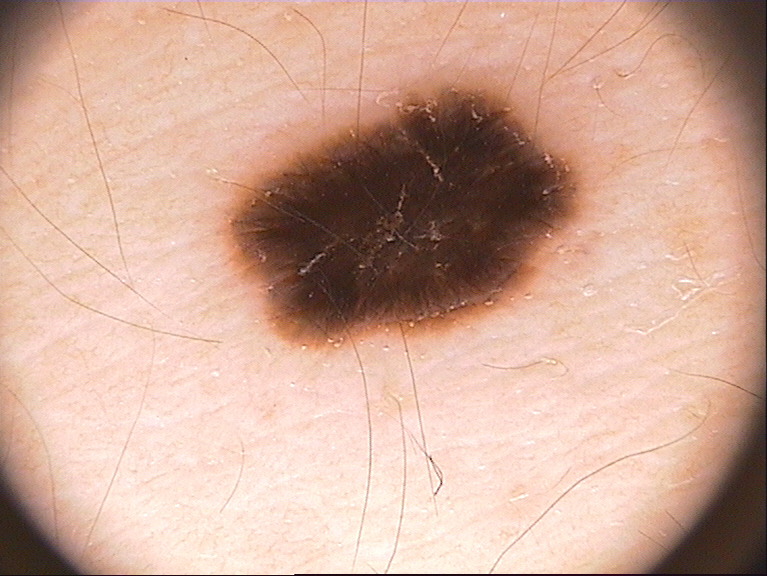}
\includegraphics[width=1.5cm,height=1.5cm]{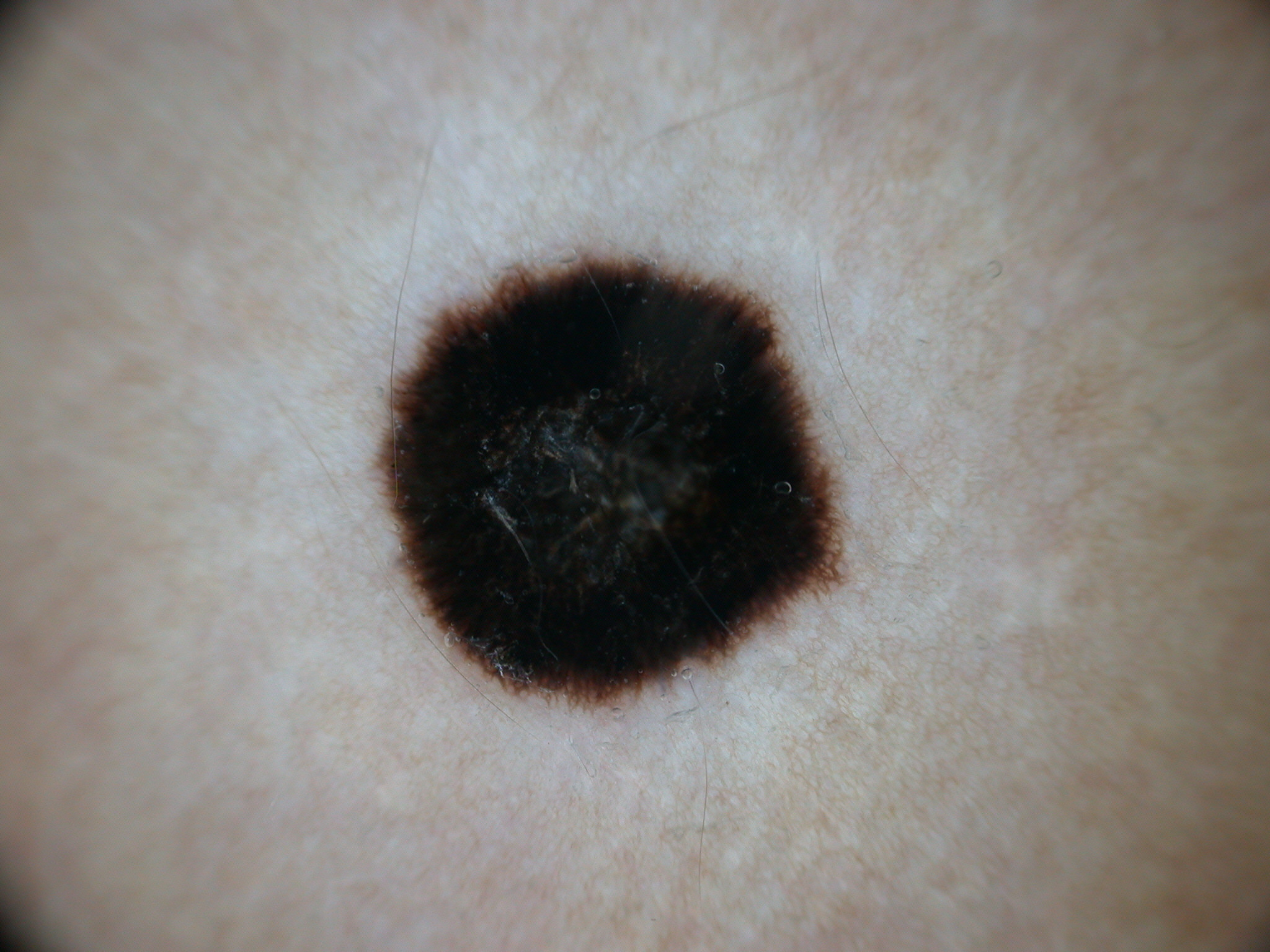}
\\\centering\text{(a)\qquad\quad(b)\quad\qquad(c)\quad\qquad(d)\qquad\quad(e)\qquad\quad(f)\qquad\quad(g)\qquad\qquad\quad}\\
\includegraphics[width=1.5cm,height=1.5cm]{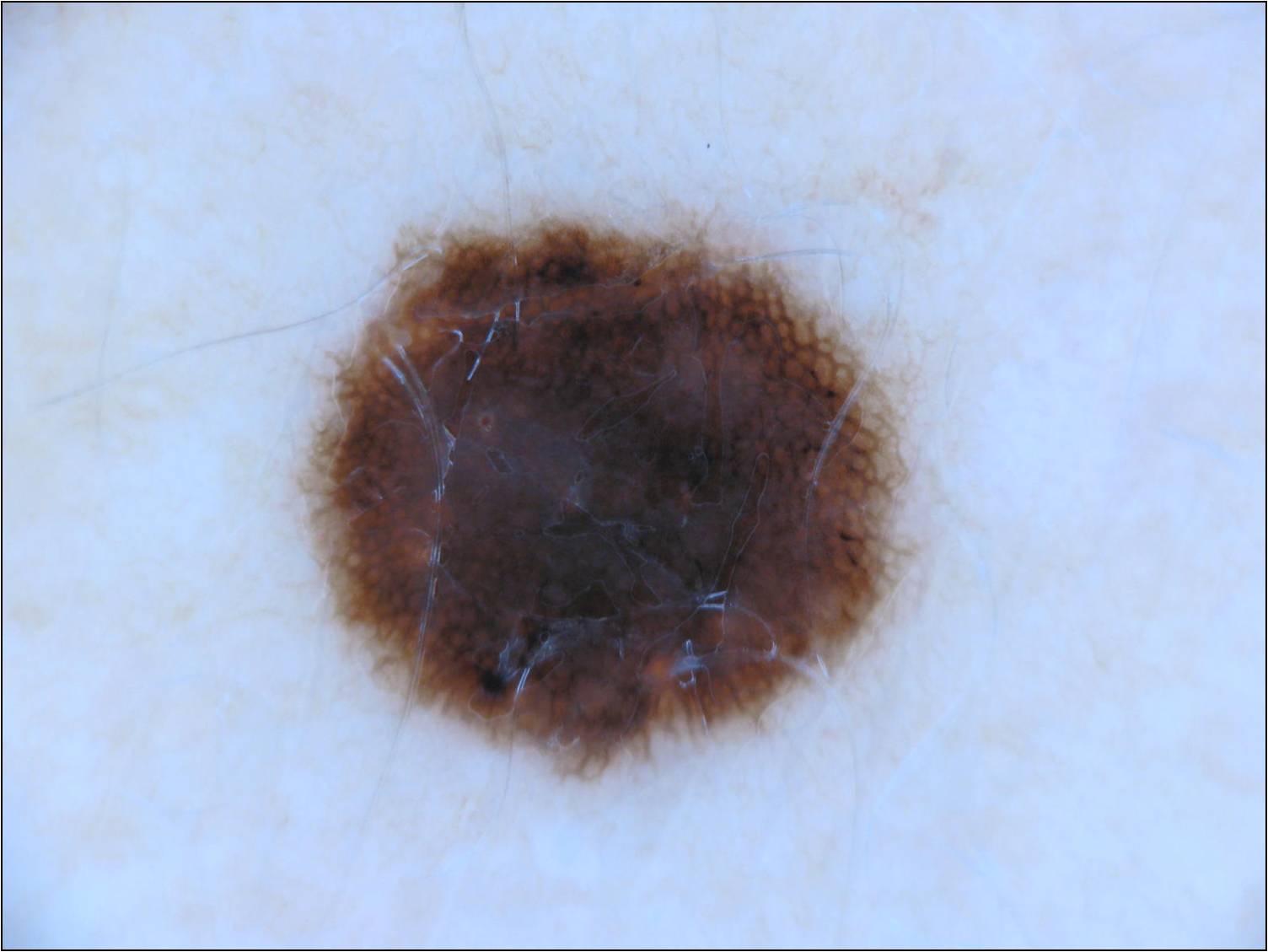}
\includegraphics[width=1.5cm,height=1.5cm]{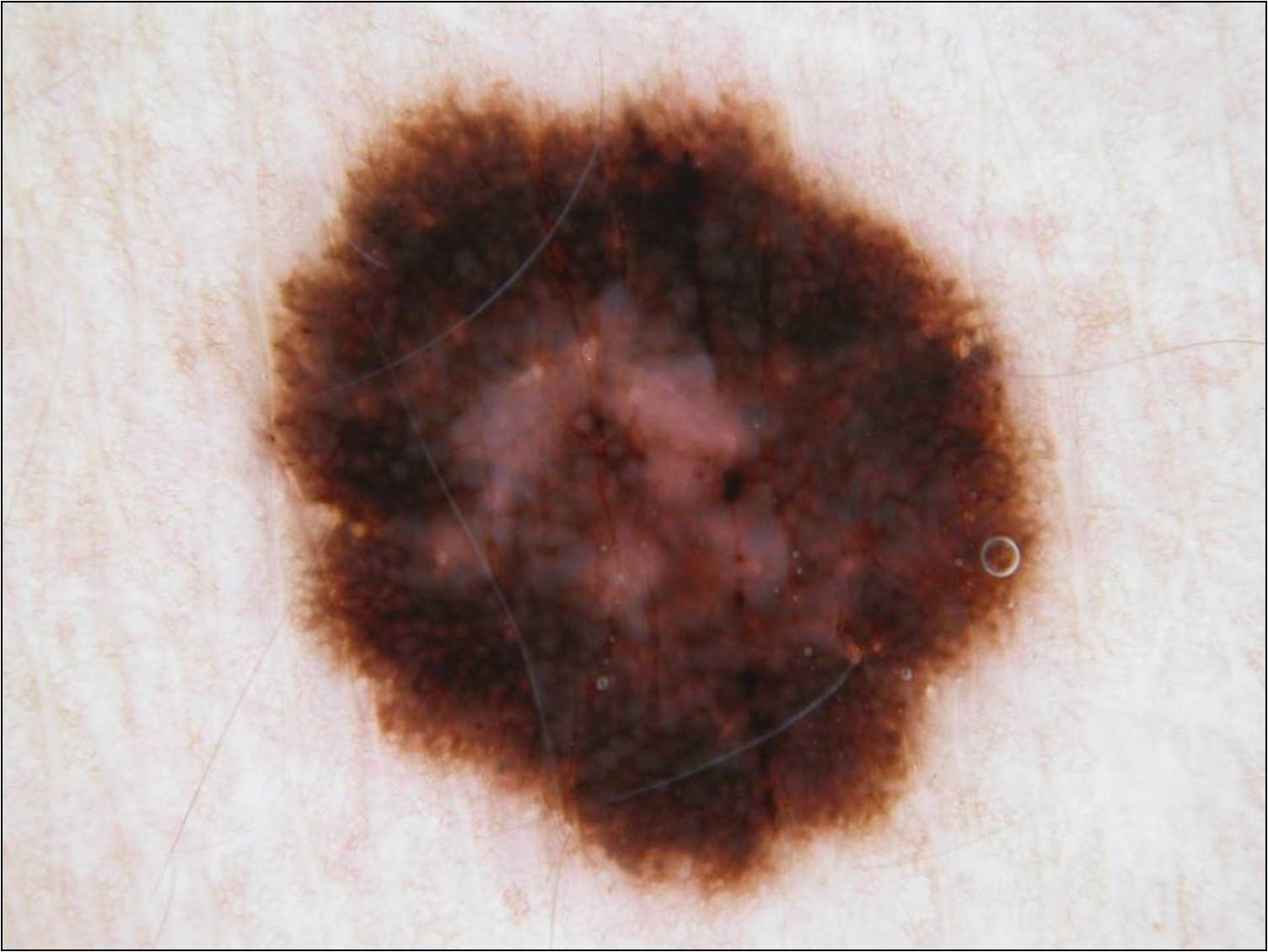}
\includegraphics[width=1.5cm,height=1.5cm]{M1.jpg}
\includegraphics[width=1.5cm,height=1.5cm]{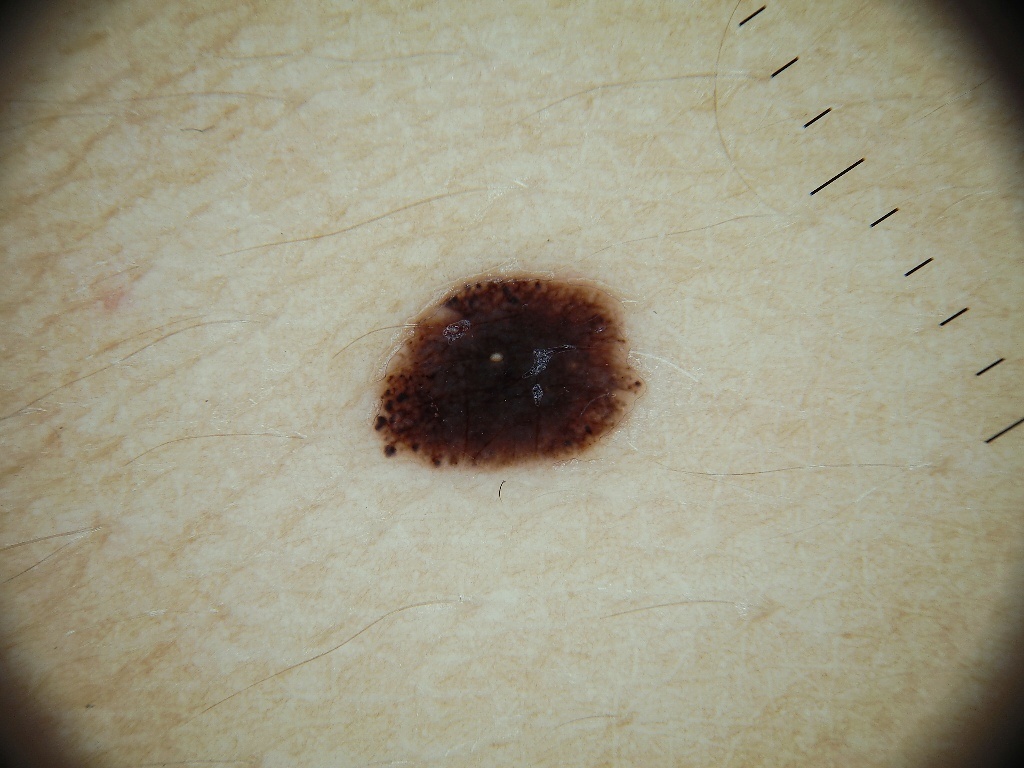}
\\\centering\text{(h)\qquad\quad(i)\qquad\quad(j)\qquad\quad(k)}
\caption{The samples with similar shape for (a)-(g) malignant  and (h)-(k) benign melanomas} 
\label{fig:Dis}
\end{figure}

The sensitivity, specificity, and accuracy of the expert's segmentation were better than those of Canny, B-Otsu, and Chan-Vese methods, but were worse than PSM and MAM, as shown in Table \ref{tab:seg900}.  Even though the expert diagnose samples one by one based on  his own segmentation, we only used the expert's segmentation followed by our own procedure to diagnose melanoma.  As shown in Fig. \ref{fig:Worse} (a) and (b), the expert's segmentation is much larger than that of PSM and MAM, but in these cases the segmentation of PSM and MAM works well to diagnose melanoma through our suggested feature extraction and thresholding methods. However, the segmentation of PSM and MAM should overcome any under- or over-segmentation due to color inhomogeneity as in Fig. \ref{fig:Worse}  (c) and (d), a hair (which is not removed by dull razor method) as in Fig. \ref{fig:Worse} (e), or a bubble  as in Fig. \ref{fig:Worse} (f). Similar-shaped skin samples are shown in Fig. \ref{fig:Dis} with malignant melanomas shown in ((a)-(g)) and benign melonams shown in ((h)-(k)). As shown in Table \ref{tab:discrimination}, the classification with MAM and PSM is better than that with the expert's segmentation. 

\begin{table}[H]
\begin{centering}
\begin{tabular}{|cllll|}
\hline
Segmentation &Accuracy &Spec. &Sens. &Jaccard\\
\hline \hline
Expert                &0.6278        &0.6105           &0.6989          &\\
\hline
Canny                &0.5300       &0.5221             &0.5625           &0.3882\\
\hline
B-Otsu               &0.5989       &0.5925           &0.6250         &0.6656\\
\hline
Chan-Vese      &0.6133         &0.6022           &0.6591          &0.6739\\
\hline
PSM                  &0.6633          &0.6464          &0.7330          & 0.7238\\
\hline
MAM                 & 0.6600         &0.6353           &0.7614           &0.7655\\
 \hline
\end{tabular}
\caption{Comparison of the classification of the five segmentation methods and the expert segmentation 
for the 900 samples. `Spec.'  and `Sens.' stand for `Specificity' and `Sensitivity', respectively.}
\label{tab:seg900}
\end{centering}
\end{table}

\begin{table}[H]
\begin{centering}
\begin{tabular}{|c|cccc|}
\hline
&G.T. & Expert & PSM & MAM\\
\hline\hline
(a)&  Benign  &{\color{red} Benign}	&{\color{red} Benign}	&{\color{red} Benign}\\
\hline
(b)&  Benign &{\color{red} Benign}&{\color{red} Benign}    &{\color{red} Benign}\\
\hline
(c)&  Benign & {\color{red} Benign}&{\color{red} Benign}	&{\color{red} Benign}\\
\hline
(d)&  Benign & {\color{red} Benign}	&{\color{red} Benign}     &{\color{red} Benign}\\
\hline
(e)& Benign &{\color{blue} Malignant} & {\color{blue} Malignant}  & {\color{red} Benign}\\
\hline
(f)&   Benign & {\color{blue} Malignant}  &{\color{blue} Malignant} 	&{\color{red} Benign}\\
\hline
(g)& Benign  &{\color{blue} Malignant} 	&{\color{blue} Malignant} 	&{\color{blue} Malignant}  \\
\hline
(h)& Malignant  &{\color{blue} Benign}	&{\color{red} Malignant}	&{\color{blue} Benign}\\
\hline
(i)& Malignant &{\color{red} Malignant}	&{\color{blue} Benign}	&{\color{red} Malignant} \\
\hline
(j)& Malignant  &{\color{blue} Benign}	&{\color{red} Malignant} 	&{\color{red} Malignant} \\
\hline
(k)& Malignant  &{\color{red} Malignant} 	&{\color{red} Malignant} 	&{\color{red} Malignant} \\
\hline\hline
Correct &  & 6 /11 & 7 /11 &9 /11\\
\hline
\end{tabular}
\caption{The diagnosis ground truth (G.T.) and classification results using the expert's, PSM, and MAM for the samples in Fig. \ref{fig:Dis}. }
\label{tab:discrimination}
\end{centering}
\end{table}

\section{Conclusions}
We proposed two segmentation methods, named PSM and MAM, in this paper. These methods are compared to three conventional methods,  
Canny, B-Otsu, and Chan-Vese, as well as an expert's segmentation using a data set taken from \cite{ISIC}. The Jacobi index of PSM and MAM, with respect to an  expert's segmentation used as the ground truth, are better than the other conventional methods over the 900 skin samples in the test data. The classification of malignant melanoma from images segmented by PSM and MAM  are better than those by the other three conventional methods as well those by the expert over the 900 skin samples. Although the segmentations made by PSM and MAM are still worse than the expert's segmentation in cases that include color inhomogeneity, hair, or bubbles, the overall efficiency of the proposed methods are still better than the expert for classification using the feature extraction and thresholding procedure suggested in this paper. 

%

\section*{Acknowledgment}

This work is supported by the Basic Research Program through the National Research Foundation of Korea  (NRF) funded by the Ministry of Science and ICT (NRF-2017R1A2B4004943).

\end{document}